\documentclass[amsmath,twocolumn,showpacs,aps,prb]{revtex4}
\usepackage{bm}
\usepackage{graphicx}
\begin{document}

\title{Attraction-repulsion transition in the interaction of adatoms and vacancies in graphene}

\author{S. LeBohec, J. Talbot, and E. G. Mishchenko}

\affiliation{Department of Physics and Astronomy, University of Utah, Salt Lake City, UT 84112, USA}

\begin{abstract}
The interaction of two resonant impurities in graphene has been predicted to have a long-range character with weaker repulsion when the two adatoms reside on the same  sublattice and stronger attraction when they are on different sublattices. We reveal that this attraction results from a single energy level. This opens up a possibility of controlling the sign of the impurity interaction via the adjustment of the chemical potential. For many randomly distributed impurities (adatoms or vacancies) this may offer a way to achieve a controlled transition from aggregation to dispersion.
\end{abstract}

 \maketitle

\section{Introduction}
Good electric conduction of intrinsic graphene \cite{CN} presents an obstacle for its use in transistor devices. The modification of graphene properties in a controllable way
is thus strongly desired, including the possibility of opening a gap. The gapless nature of the graphene spectrum, however, is protected by the equivalence of the two sublattices. This symmetry can be removed in a number of ways. Bilayer stacking breaks the equivalence of the sublattices by virtue of tunneling and allows to open the gap when an interlayer electric bias is applied \cite{CF,KCM,ZTG,MLS}. Other possibilities include breaking the symmetry by the sublattice potential\cite{GKB}, by means of the elastic strain \cite{NWM,NYL,SN,KZJ,PCP,CCC}, making finite-width nanoribbons \cite{E,SCL,HOZ}, or inducing strong spin-orbital coupling\cite{KM,QYF,WHA}. Another avenue is to utilize chemical doping with atoms or molecules that add or remove electrons from the conduction band\cite{BME,ENM,BJN} or facilitate strong inter-valley scattering\cite{DQF}. Properly understanding the consequences of the chemical doping makes it necessary to study the effective interaction between the dopants. The latter could create a variety of phases resulting from adatom ordering \cite{CSA,ASL,KCA} with major consequences for the possible applications. Such interaction is mediated by conduction electrons and is similar to the classic Casimir effect \cite{MT} in which virtual photons are responsible for the coupling. The honeycomb geometry of graphene, however, adds new features to this phenomenon.

The dependence of the inter-impurity interaction energy $W(R)$ in conventional metals displays Friedel oscillations with the period given by half the Fermi wavelength \cite{Friedel}. The amplitude of the oscillations decays as $1/r^D$, where $D$ is the dimensionality of the system \cite{LK}. In extrinsic graphene, in which the Fermi level is shifted away from the Dirac points ($k_F \ne 0$) Friedel oscillations are also present  but decay faster  than  expected in two-dimensions\cite{MF},
$\propto  \cos(2k_{\scriptscriptstyle F}R)/R^3$,  when {\it averaged} over the sublattices (see also Ref. \onlinecite{S}).

\begin{figure}[h]
\resizebox{.30\textwidth}{!}{\includegraphics{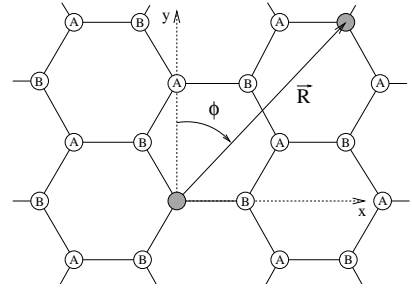}}
\caption{Carbon atoms belonging to different sublattices are labeled with $A$ and $B$. Two on-site impurities (dark circles) are placed on graphene ($AB$-configuration shown).  Periodic boundary conditions are assumed in the armchair ($x$) and zigzag ($y$) directions. The angle $\phi$ is counted from the $y$-axis.  }
\label{fig1}
\end{figure}
Additionally, the gapless character of the band spectrum of graphene allows
 to explore\cite{MilT} the ``intrinsic'' limit of $k_F\to 0$, which does not have an analog in conventional metals. When two {\it weak} on-site potential impurities of strength $U$ are present the effective interaction  depends on whether they reside on the same or different sublattices. In the former case the interaction is attractive (the derivation of Eqs.~(\ref{first_AA})-(\ref{first_AB}) is presented in the Appendices),
 \begin{equation}
 \label{first_AA}
 W_{\! AA}({\bf R})=-\frac{1}{16\pi} \frac{U^2 \! A_0^2}{v R^3}\cos^2\!{\theta_{\! AA}},
  \end{equation}
where $v$ is the graphene Dirac velocity and $A_0$ is the area of a graphene unit cell. The angle $ \theta_{\! AA}  ({\bf R}) = \frac{2\pi R}{3\! \sqrt{3}a} \cos\phi$ depends on both the length of the radius-vector ${\bf R}$ and the angle $\phi$ it makes with the zigzag direction, see Fig.~1. In the case of impurities on different sublattices the interaction is stronger and repulsive,
 \begin{equation}
 \label{first_AB}
 W_{\! AB}({\bf R})=\frac{3}{16\pi} \frac{U^2 \! A_0^2}{v R^3}\sin^2\!{\theta_{\! AB}},
  \end{equation}
 where ${\theta_{\! AB}}({\bf R})=\frac{2\pi R}{3\! \sqrt{3}a} \cos\phi+\phi$.
Eqs.~(\ref{first_AA}) and (\ref{first_AB}) can be interpreted in terms of the renormalization of the whole electron energy band in response to the presence of the impurities.

   At distances $R<UA_0/v$, the first Born approximation breaks down and the infinite series resummation taking into account multiple electron scattering off impurities has to be performed \cite{KK}. The amplitude of multiple scattering from a single impurity is given by the energy-dependent $T$-matrix, $T(E) = U/ [1-U G(0)]$, expressed via the
electron Green's function $G(0) = -\frac{EA_0}{\pi v^2}
\ln(v/a|E|)$, $a$ being the interatomic spacing.
    In the strong impurity limit, $R \ll UA_0/v$, the interaction cancels out of the scattering amplitude. This situation of a resonant impurity can also be realized with an Anderson impurity whose localized level is close to the Dirac point \cite{WKL}. Since the  strength $U$ can thus not enter the expression for the effective  energy, by dimension, it can only be given by the ratio $v/R$. In particular, when both impurities reside on the same sublattice \cite{SAL}:
   \begin{equation}
 \label{ShytovAA}
 W_{\! AA}({\bf R}) =\frac{\pi v \cos^2\!{\theta_{\! AA}}}{2R\ln^2\!{(R/a)}}.
 \end{equation}
 Notably, the interaction is repulsive, in contrast to the weak-$U$ limit, Eq.~(\ref{first_AA}).
   Similarly, the interaction between impurities residing on different sublattices also reverses sign,
 \begin{equation}
 \label{ShytovAB}
 W_{\! AB}({\bf R}) =-\frac{2v |\sin{\theta_{\! AB}}|} {R\ln{(R/a)}} +\frac{\pi v \sin^2\!{\theta_{\! AB}}}{2R\ln^2\!{(R/a)}},
 \end{equation}
 The first term in Eq.~(\ref{ShytovAB}), derived in Ref.~\onlinecite{SAL}, dominates (when $\ln(R/a) \gg 1$) over the second (repulsive) term, whose derivation is given in Appendix B.
Both $W_{\! AA}$ and the second term in $W_{\! AB}$ can be viewed as the {\it perturbative}  renormalization of the continuous spectrum to the lowest order in the effective impurity strength $U_{eff}=T(E) \sim v^2/[EA_0 \ln(v/a|E|)] \sim vR /[A_0\ln{(R/a)]}$, since the relevant energies are $E\sim v/R$. Substituting this expression in place of $U$ in Eq.~(\ref{first_AB}), we recover the right estimate of the effect. This is not surprising since, as explained above, the actual dimensionless parameter that controls the effective strength of the impurity is $U_{eff}\! A_0/vR\sim \ln^{-1}{(R/a)}\ll 1$.

By contrast, the leading attractive term in Eq.~(\ref{ShytovAB}) is {\it non-perturbative}.
 We are now going to demonstrate  that a {\it single} impurity level is responsible for this contribution to $W_{\! AB}$, and explain that this understanding leads to the possibility of controlling the sign of the interaction by adjusting the chemical potential.
 The sensitivity of the interaction between two adatom to the chemical potential has previously been reported on the basis of numerical studies \cite{SJR}, however, the underlying physical mechanism of the impurity level formation has not been elucidated nor has it been shown to extend to the case of many randomly distributed adatoms as is the subject of this paper.

\section{Energy levels of two impurities}

 We consider a tight-binding model of $\pi$-electrons in graphene interacting with two on-site potential impurities positioned at ${\bf r}=0$ and ${\bf r}={\bf R}$, see Fig.~\ref{fig1},
\begin{eqnarray}
\label{hamilton}
\hat H=t\sum_{{\bf r}_A}\sum_{i=1,2,3} \hat a^\dagger ({\bf r}_A) \hat b({\bf r}_A+{\bm a}_i) +\text{h.c.} \nonumber\\
+U \hat a^\dagger(0) \hat a(0) +U \hat b^\dagger ({\bf R}) \hat b ({\bf R}).
\end{eqnarray}
The operators $\hat a^\dagger$($\hat b^\dagger$) create electrons on the corresponding sites of the sublattice $A$($B$); the vectors ${\bm a}_i$ connect $A$-atoms with their three nearest $B$-neighbors. The Hamiltonian (\ref{hamilton}) is written for the case of the second impurity residing on the sublattice $B$ (otherwise the operators $\hat b$ have to be replaced with $\hat a$ in the last term).  From the Hamiltonian (\ref{hamilton}) in the Fourier representation with $\hat a({\bf r}_A)=\sqrt\frac{2}{N}\sum_{\bf k} \hat a({\bf k}) e^{i{\bf k} {\bf r}_{A}-iE t}$, we find the following equations of motion for the electron operators,
\begin{eqnarray}
E \hat a({\bf k})&=&t({\bf k}) \hat b({\bf k})+\frac{2U}{N} \sum_{\bf k'} \hat a({\bf k'}),\\
E \hat b({\bf k})&=&t^*({\bf k}) \hat a({\bf k})+\frac{2U}{N} \sum_{\bf k'} \hat b({\bf k'})e^{i({\bf k'}-{\bf k})  {\bf R}},
\end{eqnarray}
where $t({\bf k})=t\sum_{i} e^{i{\bf k}{\bm a}_i}$ and $N$ is the total number of carbon atoms.
The solution of these equations is straightforward and yields the following condition for the energy spectrum of the two-impurity $AB$-configuration,
\begin{equation}
\label{AB}
\Bigl[1-U\! \sum_{\bf k}A({\bf k},0)\Bigr]^2 = U^2\! \sum_{\bf k}B({\bf k},{\bf R})\! \sum_{{\bf k}'}B(-{\bf k}',{\bf R}),
\end{equation}
where
\begin{equation*}
\left\{\begin{array}{l} A({\bf k},{\bf R}) \\ B({\bf k},{\bf R})\end{array} \right\} =
\frac{2}{N} \frac{e^{-i{\bf k} {\bf R}}}{(E+i\eta)^2-|t({\bf k})|^2} \left\{\begin{array}{c} E \\ t({\bf k})\end{array} \right\},
\end{equation*}
Similarly, for the $AA$-configuration,
\begin{equation}
\label{AA}
\Bigl[1-U\! \sum_{\bf k}A({\bf k},0)\Bigr]^2 = U^2\! \sum_{\bf k}A({\bf k},{\bf R})\! \sum_{{\bf k}'}A(-{\bf k}',{\bf R}).
\end{equation}
The integrals over the quasimomentum ${\bf k}$ are taken over the hexagonal Brillouin zone. In the low energy sector only the vicinities of the two Dirac points determined from the condition $t({\bf K}_{\pm})=0$: ${\bf K}_{\pm}=\frac{2\pi}{3a}(1,\pm 1/\sqrt{3})$ are important.  Up to an irrelevant common phase factor, $t({\bf k})\approx v (q_x\pm i q_y),$ where ${\bf q}={\bf k}-{\bf K}_{\pm}$.

{\it $AB$-configuration.} Performing the integrals in Eq.~(\ref{AB}) we obtain the dispersion equation in the form,
\begin{eqnarray}
\label{AB_dispersion}
\left(1+\frac{UA_0 E}{\pi v^2}\Bigl[\ln{\left| t/E\right|}+i\pi/2\Bigr]\right)^2= - \frac{U^2A^2_0}{4v^4}  \nonumber\\
\times \sin^2\theta_{\! AB} E^2 \left[ H_1^{(1)}\left(\frac{ER}{v} \right)\right]^2,
\end{eqnarray}
where $A_0= 3\sqrt{3} a^2/2$ is the area of a unit cell.
In the logarithmic approximation, the Hankel function can be replaced with its value for small arguments, $H_1^{(1)}(x)\approx -2i/(\pi x)$, yielding the impurity levels,
\begin{equation}
\label{ABlevels}
E_{\! AB}=\frac{(\pm U-U_c) v |\sin\theta_{\! AB}|} {UR[\ln{(R/a)}+i\pi/2]},~~~ U_c=\frac{\pi v R }{A_0|\sin\theta_{\! AB}|}.
\end{equation}
Due to the overlap with the continuum of propagating states the levels have finite width, which is small by $\ln^{-1}{(R/a)} \ll 1$.
Above the ``critical'' value $U_c$  the higher of the two levels crosses over the Fermi level $\mu=0$ and becomes depopulated. In the limit of $U\gg U_c$ the two levels become symmetric with respect to $E=0$. When the chemical potential is $\mu=0$, the energy of the lower (filled) level exactly reproduces the leading attraction term in Eq.~(\ref{ShytovAB}), if spin degeneracy is taken into account. It is this ``Dirac point crossing'' that is responsible for the attraction in ${\it AB}$ case, see Fig.~2 (left).

\begin{figure}[h]
\resizebox{.48\textwidth}{!}{\includegraphics{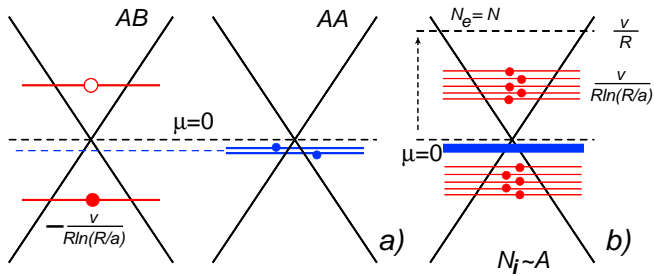}}
\caption{The energy spectrum of graphene with resonant impurities. a) Two impurities: in the $AB$-configuration a single-impurity level is split into two, $ \pm \frac{v|\sin\theta_{\! AB}|}{R\ln{(R/a)}}$. For a fixed chemical potential $\mu=0$, the upper level is empty, while the dependence of the lower level on the inter-impurity distance reproduces the attraction term in Eq.~(\ref{ShytovAB}); in the $AA$-configuration both the splitting and the energies are negligible, the impurity levels stay on the same side of the Dirac point (below the chemical potential for $U>0$). The interaction comes from the renormalization of the energies of propagating states and is repulsive, Eq.~(\ref{ShytovAA}). b) When the number of impurities $N_i$ scales with the size $A$ of the system the sign of the interaction can change: when $\mu =0$ the attraction from the negative energy states dominates; if the number of electrons is fixed instead, $N_e=N$, the states up to the chemical potential $\sim v/R$ are  populated, thus negating the effects of the negative energy states and  leading to the repulsion due to the renormalization of the propagating states. }
\label{fig2}
\end{figure}

{\it $AA$-configuration.} The purely repulsive character of the interaction of two impurities residing on the same sublattice can be traced to a completely different behavior of the impurity levels. Calculating the integral in the right hand side of Eq.~(\ref{AA}), we obtain an equation similar to Eq.~(\ref{AB_dispersion}) with the following changes: $H_1^{(1)} \to H_0^{(1)}$ and $\sin\theta_{\! AB} \to \cos\theta_{\! AA}$. Since $H_0^{(1)}$ is only logarithmically divergent at small arguments, the solutions with $E=0$ are {\it absent}. This means that no impurity state can cross the Dirac point. For large values of $U\gg t$, both values are very close to the $E=0$ level, $E_{\! AA} \sim  -t^2/U \ln{(t/U)}$, and contribute negligibly to the total energy of the system, as illustrated in Fig. 2 (center). The interaction energy in the strong-$U$ limit, thus, is entirely due to the renormalization of the band spectrum, Eq.~(\ref{ShytovAA}).

\section{Attraction-repulsion transition}

 \subsection{Two AB impurities}

 The chemical potential $\mu$ can be controlled by means of electrostatics via leads and/or gates. Decreasing $\mu$ below the energy of the lower impurity state, Eq.~(\ref{ABlevels}), or increasing it above the upper level (so that both levels are empty or populated) would negate the effects of the impurity levels and lead to the disappearance of the attractive contribution in Eq.~(\ref{ShytovAB}) rendering the residual interaction {\it repulsive}. Let us emphasize that this sign reversal is different from the Friedel oscillations in a doped graphene. The latter develop when $k_F R \sim 1$ while in our case significantly lower changes in the chemical potential are needed, $k_F R \sim \ln^{-1}(R/a)\ll 1$. We are now going to show that this effect survives when the number of impurities scales with the size of the system.

 \subsection{Many randomly distributed impurities}
 When impurities with finite density $N_i/A$ are randomly distributed in the system, the stronger attraction from $AB$ pairs dominates over the weaker repulsion of $AA$ and $BB$ pairs  \cite{SAL}. Our numerical findings also support this for $\mu=0$, but with increasing $\mu$, the transition to the repulsive regime occurs, similarly to the two-impurity case.
In particular, we considered rectangular graphene samples described by the Hamiltonian (\ref{hamilton}) with $n_x \times n_y$  atoms, and\cite{shytov} $U=100t$. Periodic boundary conditions are imposed along both armchair ($x$) and zigzag ($y$) directions.
The energy spectrum of the sample is found by the exact diagonalization of the Hamiltonian and the sum over all filled states is then taken to give the total energy $E_{N_i}$ in the presence of $N_i$ impurities.
The interaction energy $W_{N_i}$ is obtained by subtracting the energy of independent impurities,
\begin{equation}
\label{interaction_energy_definition}
W_{N_i}=E_{N_i}-E_0-N_i \left(E_1-E_0\right).
\end{equation}
The definition (\ref{interaction_energy_definition}) is different from that of Ref.~\onlinecite{SAL}, where the term linear in $N_i$ was allowed to be an adjustable fitting parameter. 

While we are in a qualitative agreement with Ref.~\cite{SAL} in case of $\mu=0$, our numerical results differ significantly from those reported in Ref.~\onlinecite{SAL} when {\it the number of electrons $N_e$} corresponds to $\mu \ne 0$. 
Most notably, we obtain that the sign of $W_{N_i}$ can be reversed if the chemical potential is set sufficiently high, see Fig.~3. This would occur, for example, if the impurities were placed on an isolated sheet of graphene so that $N_e$ is kept equal to the total number of carbon sites $N$. The sign reversal can be explained with the help of the same ``Dirac point crossing'' picture illustrated in Fig.~\ref{fig2}. When $N_i$ resonant impurities are spread over the system, the same number of low-energy levels are created. Since impurities are distributed randomly and uniformly over  $A$ and $B$ sites, two phenomena occur simultaneously: ``{\it AA}-type'' accumulation just below $E=0$  within a narrow energy range $\propto U^{-1}$, and the formation of the ``{\it AB}-type'' impurity bands on both sides of $E=0$. The number of states that cross the $E=0$ level is $\alpha N_i$, with $(1-2 \alpha) N_i$ levels accumulated near $E=0$  (for small concentrations $N_i/A$, we observe that $\alpha \approx 0.43$). We stress that the attraction of the impurities ($W_{N_i}<0$) is {\it solely} due to the fact that those states that crossed the Dirac point remain {\it unfilled} when $\mu=0$.

\begin{figure}[h]
\resizebox{.48\textwidth}{!}{\includegraphics{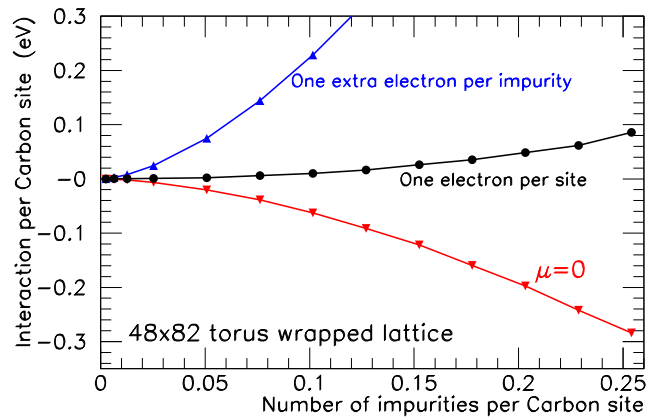}}
\caption{ The interaction energy $W_{N_i}/N$ per site as a function of the impurity density $N_i/N$ on a torus-wrapped ($N=48\times 82$) graphene lattice. The lower curve represents the fixed chemical potential, $\mu=0$, case. The middle curve corresponds to $N_e=N$ (number of electrons fixed at doping) and the top one stands for one extra electron per impurity, $N_e=N+N_i$. Each point represent the average of 20 numerical experiments. The standard deviation is smaller than the points. }
\label{fig3}
\end{figure}

To the contrary, if, for example, the number of electrons is kept fixed instead ($N_e =N$), exactly $\alpha N_{i}$ levels with positive energies have to be occupied. Occupation of each impurity state is detrimental to the attraction. Not all of the impurity states, however, are going to be occupied as other (propagating) states of similar energies ``compete'' for the same $2\alpha N_{i}$ electrons (taking into account spin degeneracy). Nevertheless, it is easy to see that, in the logarithmic approximation, $\ln{(R/a)}\gg 1$, {\it most} of the impurity states will be occupied. Indeed, one would require the linear band to be filled up to the energy $\sim v\sqrt{\frac{\alpha N_i}{A}}$ to accommodate $2\alpha N_i$ electrons. On the other hand, the characteristic energy scale of the impurity band is smaller, $\frac{v}{ R} \ln^{-1}{( R/a)} \sim  v \ln^{-1}{( R/a)}\sqrt{\frac{ N_i}{A}}$. In other words, the mean level spacing in the impurity band is logarithmically smaller than the level spacing of the propagating states, resulting in the large fraction of  the former being populated when $N_e$ is made equal  to the number of carbon sites $N$ or exceed it, which is the case when $\mu$ is increased further. Fig.~\ref{fig3} provides a numerical confirmation of these semi-qualitative arguments. Additionally, we numerically observe that the ratio of the number of electrons that need to be removed from the $\pi$-band to the total number of impurity atoms to reach the attraction to repulsion transition is $\sim15\%$ rather independently from the concentration of impurities in the range from $5\%$ to $35\%$.
 
\subsection{Vacancies}

Another realization of a resonant impurity limit is the case of a missing carbon atom. We observe numerically that modeling such vacancies in terms of zero hoppings to/from the neighboring sites gives results that are very close to the  model of a strong on-site potential $U$. Another difference is that in a neutral graphene with $N_i$ vacancies $N_i$ electrons are missing from the $\pi$-band. Since the number of states ``escaped'' through the Dirac point is $2\alpha N_i$ (twice the number of levels), which is somewhat less than $N_i$, the chemical potential of the neutral graphene with vacancies is {\it negative} (but close to $E=0$), resulting in the attraction of vacancies. This is opposite to the sign of the interaction in a neutral sample with potential impurities (where $N_e=N$). Still, when the interaction is studied as a function of the chemical potential the two cases yield virtually indistinguishable results. For this reason datasets for vacancies are not shown in Fig.~\ref{fig3}.

 \section{Summary}

 The interaction of resonant impurities in graphene displays a transition in their net interaction from attraction to repulsion depending on the chemical potential. This phenomenon is traced to the existence of impurity levels with energies $E\sim \pm v/R$ that appear when impurities reside on the opposite sublattices. Asymmetric filling of such states, which occurs for a chemical potential close to the Dirac point $E=0$, favors attraction. With the change of the chemical potential the interaction becomes repulsive as the continuum of propagating states dominates. This mechanism suggests the possibility of a transition from aggregation of adatoms to their dispersion, which could be advantageous for graphene functionalization. In particular, the possibility of controlling the conduction properties of graphene can be envisaged, with the metallic phase realized when gap-opening adatoms are aggregated in a small area of a graphene device and the semiconducting state occurring when they are spread uniformly across its entire extent. Similarly, a ``nanobreaker'' could be realized with the help of vacancies, whose aggregation will result in the loss of mechanical stability of the graphene sheet.

\section{Acknowledgements}

We thank D. Pesin, M.~Raikh and O.~Starykh
for fruitful discussions. S.L. and J.T. are supported by NSF
through MRSEC DMR-1121252; E.M. acknowledges support from the Department of Energy,
Office of Basic Energy Sciences, Grant No.~DE-FG02-06ER46313.

\appendix

\section{Interaction energy of two on-site impurities}

It is convenient to express the interaction energy  $W({\bf R})$ of two on-site impurities
placed a distance ${\bf R}$ away from each other via the electron Green's function (found in Appendix B).
We start with the Hamiltonian of the two impurities:
\begin{equation}
\label{hamiltonian} H=t\sum_{{\bf r}}\sum_{i=1,2,3} \hat c^\dagger ({\bf r}) \hat c({\bf r}+{\bm a}_i) + U\hat c^\dagger (0)
\hat c (0)+U\hat c^\dagger ({\bf R}) c({\bf R}).
\end{equation}
Here $\hat c({\bf r})=\hat a ({\bf r})$ when ${\bf r}$ belongs to sublattice {\it A} and $\hat c({\bf r})=\hat b ({\bf r})$ when it belongs to {\it B}. The summation over ${\bf r}$ in Eq.~(\ref{hamiltonian})  is taken over {\it both} sublattices, in order to cast the Hamiltonian (5) in a more compact form. The interaction energy is most simply found from the following identity\cite{LLQM},
\begin{equation}
\label{LLidentity} \frac{\partial W}{\partial U}= \left\langle
\frac{\partial H}{\partial U}\right\rangle= -i {\cal G}(0,0,t=-0)-i{\cal G} ({\bf R},{\bf R},t=-0).
\end{equation}
Here Green's function is determined in the usual way,
\begin{equation}
\label{greensfunction} {\cal G}({\bf r},{\bf r'},t)
=-i\langle T \hat c({\bf r},t)\hat c^\dagger({\bf
r'},0)\rangle.
\end{equation}
The interaction energy is therefore
\begin{equation}
\label{potenergy} W({\bf R})=-2i \int\limits_0^U dU~ \Bigl[
{\cal G}(0,0,t=-0)+{\cal G}({\bf R},{\bf R},t=-0)\Bigr],
\end{equation}
here the factor $2$ takes into account spin degeneracy.
The problem is thus reduced to finding the Green's function in the
presence of two impurities.

\section{Green's function of the two-impurity problem}

From the equations of motion for the electron operators $i\partial \hat c({\bf r},t)/\partial t =[\hat c({\bf r},t),H]$ the equation for the Green's function $\hat {\cal G}_E({\bf r},{\bf
r'})$ in the energy representation is found:
\begin{eqnarray}
\label{greens_equation} E {\cal G}_E({\bf r},{\bf
r'})- t\sum_{i} {\cal G}_E({\bf r}+{\bm a}_i,{\bf
r'})-U \delta_{{\bf r},0} {\cal G}_E(0,{\bf
r'})\nonumber\\ -U\delta_{{\bf r},{\bf R}} {\cal G}_E({\bf R},{\bf
r'})=\delta_{{\bf r},{\bf r'}}.
\end{eqnarray}
We look for a solution of
Eq.~(\ref{greens_equation}) in the form,
\begin{equation}
\label{ansatz} {\cal G}_E({\bf r},{\bf r'})= G_E({\bf r},{\bf
r'})+ G_E({\bf r},0) A ({\bf r'})+ G_E({\bf r},{\bf R}) B({\bf r'}),
\end{equation}
where $ G_E({\bf r},{\bf r'})$ is the Green's function of the free electrons
in graphene. Substituting Eq.~(\ref{ansatz}) into the equation
(\ref{greens_equation}) we find two equations for the functions
$A({\bf r})$ and $ B({\bf r})$,
\begin{eqnarray}
\label{ABeq}
 A ({\bf r'})-T_E   G_E(0,{\bf R}) B ({\bf
r'})&=&T_E  G_E(0,{\bf r'}),\nonumber\\
-T_E G_E({\bf R},0) A ({\bf r'})+B ({\bf r'})&=&T_E
G_E({\bf R},{\bf r'}),
\end{eqnarray}
here the $T_E$-matrix is introduced,
\begin{equation}
\label{Tmatrixdefinition} T_E=\frac{U}{1-U G_E(0,0)}.
\end{equation}
Solutions of Eqs.~(\ref{ABeq}) are (argument $E$ dropped for brevity)
\begin{eqnarray}
&& A ({\bf r'})=T \frac{ G(0,{\bf r'})+T G(0,{\bf R})
G({\bf R},{\bf r'})}{1-T^2 G(0,{\bf R}) G({\bf
R},0)} \nonumber\\
&& B ({\bf r'})=T\frac{G({\bf R},{\bf r'})+T G({\bf
R},0) G(0,{\bf r'})}{1-T^2 G(0,{\bf R}) G({\bf
R},0)}.
\end{eqnarray}
Substituting these expressions into Eq.~(\ref{ansatz}), we obtain
\begin{widetext}
\begin{equation}
\label{sum}  {\cal G}(0,0)+{\cal G}({\bf R},{\bf
R})=2G(0,0)+\frac{2T[G^2(0,0) + G({\bf R},0) G(0,{\bf
R})]+4T^2 G(0,0) G({\bf R},0) G(0,{\bf R})}{1-T^2 G(0,{\bf R}) G({\bf
R},0)}.
\end{equation}
\end{widetext}
It is now convenient to express $T_E$ back via $U$. After simple algebra, we find that the right-hand side of Eq.~(\ref{sum}) is equal to
$$
-\frac{d}{dU}\ln{\left([1-UG(0)]^2-U^2 G({\bf R},0) G(0,{\bf
R}) \right)}.
$$
Finally, substituting this into (\ref{potenergy}), subtracting the
same expression when the two impurities are far away from each
other, ${R}\to \infty$, we obtain
\begin{equation}
\label{adatomW} W({\bf R})= 2i
\int\limits_{-\infty}^\infty\frac{dE}{2\pi}
\ln{\left(1-T^2_E G_E({\bf R},0) G_E(0,{\bf R}) \right)}.
\end{equation}
It is now convenient to make use of the fact that the time-ordered Green's functions do not have singularities in the first and third quadrants of the complex $E$-plane, and rotate the integration path counterclockwise by the angle $\pi/2$ so that it coincides with the imaginary axis, $E=i\omega$.  As a result we obtain,
\begin{equation}
\label{adatomWimag} W({\bf R})= -2
\int\limits_{-\infty}^\infty\frac{d\omega}{2\pi}
\ln{\left(1-T^2(i\omega) G_{i\omega}({\bf R},0) G_{i\omega} (0,{\bf R}) \right)}.
\end{equation}
\begin{figure}[h]
\resizebox{.48\textwidth}{!}{\includegraphics{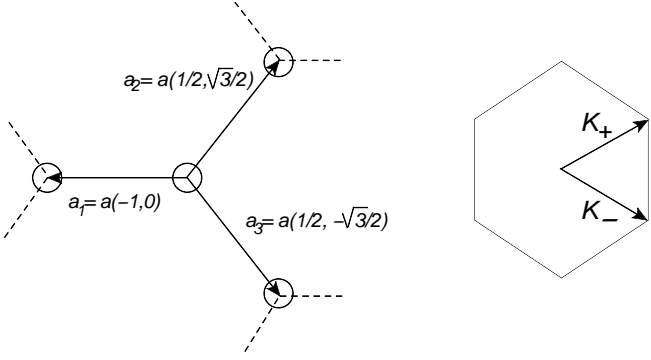}}
\caption{Graphene lattice and the first Brillouin zone, ${\bf K}_{\pm}=\frac{2\pi}{3a}(1,\pm 1/\sqrt{3})$. }
\label{fig4}
\end{figure}

The calculation of the interaction energy is now reduced to finding the free electron's Green's functions.
The same-sublattice Green's function,
\begin{equation}
\label{green_AA}
G({\bf r}_A,{\bf r}_A', t)= -i\langle T \hat a({\bf r}_A,t)\hat a^\dagger({\bf
r}_A',0)\rangle,
\end{equation}
is found with the help of the Fourier representation,
\begin{equation}
\label{eigenstates}
\hat a({\bf r}_A,t)=\sqrt\frac{2}{N}\sum_{\bf k} \sum_{\beta=\pm 1}\hat a ({\bf k}) e^{i{\bf k} {\bf r}_{A}-i\beta|t({\bf k})|t},
\end{equation}
where $\beta=1$ stands for the states above the Dirac points and $\beta=-1$ for the states below them. With $N$ being the number of carbon atoms in the system, the total number of different quasimomenta states is $N/2$. The summation is taken over the hexagonal Brillouin zone. From Eqs.~(\ref{green_AA}) and (\ref{eigenstates}) we find,
\begin{equation}
G_{i\omega}({\bf r}_A,{\bf r}_A')=-A_0 \int\frac{d^2k}{(2\pi)^2} e^{i {\bf k}({\bf r}_A-{\bf r}_A')} \frac{i\omega}{\omega^2+|t({\bf k})|^2},
\end{equation}
where $A_0$ is the area of a unit cell in a honeycomb lattice (note that $\frac{2}{N}\sum_{\bf k}$ is replaced with $A_0 \int\frac{d^2k}{(2\pi)^2}$).  For large distances $|{\bf r}_A-{\bf r}_A'|\gg a$ only the vicinities of the two Dirac points are important, ${\bf k}={\bf K}_{\pm}+{\bf q}$, determined from the condition $t({\bf K}_{\pm})=0$: ${\bf K}_{\pm}=\frac{2\pi}{3a}(1,\pm 1/\sqrt{3})$. We thus obtain,
\begin{eqnarray}
\label{G_AA}
G_{i\omega}({\bf r}_A,{\bf r}_A')&=&-i\omega A_0 \left(e^{i {\bf K}_+({\bf r}_A-{\bf r}_A')}+e^{i {\bf K}_-({\bf r}_A-{\bf r}_A')}  \right) \nonumber\\
&& \times \int\frac{d^2q}{(2\pi)^2} \frac{e^{i {\bf q}({\bf r}_A-{\bf r}_A')} }{\omega^2+v^2q^2}\nonumber\\
&=&-\frac{i\omega A_0}{2\pi v^2}  \left(e^{i {\bf K}_+({\bf r}_A-{\bf r}_A')}+e^{i {\bf K}_-({\bf r}_A-{\bf r}_A')} \right) \nonumber\\ && \times
K_0\left(\frac{|\omega||{\bf r}_A-{\bf r}_A'|}{v}\right).
\end{eqnarray}
Obviously, the Green's function $G_{i\omega}({\bf r}_B,{\bf r}_B')$ is given by the same expression. In particular, for coinciding points,
\begin{equation}
\label{G_00}
G_{i\omega}(0,0)=-\frac{i\omega A_0}{\pi v^2} \ln{\left(\frac{t}{|\omega|}\right)}.
\end{equation}
To find the function $G({\bf r}_B,{\bf r}_A',t)= -i\langle T \hat b ({\bf r}_B,t)\hat a^\dagger({\bf
r}_A',0)\rangle $, we use the identity $\hat b({\bf k})=\beta \frac{t^*({\bf k})}{|t({\bf k})|}  \hat a({\bf k})$, see Eqs.~(5-6) of the paper, which give
\begin{equation}
\label{eigenstates_b}
\hat b({\bf r}_B,t)=\sqrt\frac{2}{N}\sum_{\bf k} \sum_{\beta=\pm 1}\beta \frac{t^*({\bf k})}{|t({\bf k})|}  \hat a ({\bf k}) e^{i{\bf k} {\bf r}_{B}-i\beta|t({\bf k})|t}.
\end{equation}
As a result we arrive at
\begin{equation}
G_{i\omega}({\bf r}_B,{\bf r}_A')=-A_0 \int\frac{d^2k}{(2\pi)^2} e^{i {\bf k}({\bf r}_B-{\bf r}_A')} \frac{t^*({\bf k})}{\omega^2+|t({\bf k})|^2}.
\end{equation}
Given the choice of the vectors ${\bf a}_i$ as shown in the Fig.~\ref{fig4}, $t({\bf k})=t\sum_i e^{i{\bf k a}_i}= t[e^{-ik_xa}+2e^{ik_x a/2}\cos{(\sqrt{3}k_ya/2)}]$.
Again expanding near the two Dirac points, $t({\bf k})=t({\bf K}_\pm +{\bf q})\approx ie^{i\pi/3} v(q_x\pm i q_y)$. Upon taking the $d^2q$ integral we obtain,
\begin{widetext}
\begin{eqnarray}
G_{i\omega}({\bf r}_B,{\bf r}_A')&=&-e^{i\pi/6} A_0 \left(e^{i {\bf K}_+({\bf r}_B-{\bf r}_A')+i\phi}-e^{i {\bf K}_-({\bf r}_B-{\bf r}_A')-i\phi}  \right) \int\limits_0^\infty\frac{q^2 dq}{2\pi} \frac{vJ_1(q|{\bf r}_B-{\bf r}_A'|)} {\omega^2+v^2q^2}\nonumber\\
&=&-e^{i\pi/6} \frac{|\omega| A_0}{2\pi v^2} \left(e^{i {\bf K}_+({\bf r}_B-{\bf r}_A')+i\phi}-e^{i {\bf K}_-({\bf r}_B-{\bf r}_A')-i\phi}  \right) K_1\left(\frac{|\omega||{\bf r}_B-{\bf r}_A'|}{v}\right),
\end{eqnarray}
where the angle $\phi$ is the one vector ${\bf r}-{\bf r}'$ makes with the $y$-axis (zigzag direction).
Similarly for $G({\bf r}_A,{\bf r}_B',t)=-i\langle T \hat a ({\bf r}_A,t)\hat b^\dagger({\bf
r}_B',0)\rangle$ we obtain,
\begin{eqnarray}
G_{i\omega}({\bf r}_A,{\bf r}_B')&=&-A_0 \int\frac{d^2k}{(2\pi)^2} e^{i {\bf k}({\bf r}_A-{\bf r}_B')} \frac{t({\bf k})}{\omega^2+|t({\bf k})|^2} \nonumber\\
&=& \left(e^{i {\bf K}_+({\bf r}_A-{\bf r}_B')-i\phi}-e^{i {\bf K}_-({\bf r}_A-{\bf r}_B')+i\phi}  \right)  \times e^{-i\pi/6} \frac{|\omega| A_0}{2\pi v^2} K_1\left(\frac{|\omega||{\bf r}_A-{\bf r}_B'|}{v}\right).~~~~~~~~
\end{eqnarray}
\end{widetext}

\section{{\it AA}-configuration of two impurities}

When both impurities are residing on the same sublattice, using Eq.~(\ref{G_AA}) we find that the interaction energy is given by
\begin{eqnarray}
\label{W_AA} W_{\it \! AA}({\bf R})= -2\!\!
\int\limits_{-\infty}^\infty\! \frac{d\omega}{2\pi}
\ln{\left(1-T^2_{i\omega}G_{i\omega}({\bf R},0)G_{i\omega}(0,{\bf R})  \right)}\nonumber\\ =-2\!\!
\int\limits_{-\infty}^\infty\! \frac{d\omega}{2\pi}
\ln{\left(1+T^2_{i\omega} \frac{\omega^2 A_0^2}{\pi^2 v^4} K^2_0\left({|\omega| R}/{v}\right) \cos^2\theta_{\it \! AA} \right)},~
\end{eqnarray}
where the $T$-matrix is written with the help of Eqs.~(\ref{Tmatrixdefinition}) and (\ref{G_00}) as
\begin{equation}
\label{T-matrix}
T_{i\omega} =\frac{U}{1+\frac{i\omega U A_0}{\pi v^2} \ln{\left(\frac{t}{|\omega|}\right)}}.
\end{equation}
In the weak impurity limit $UA_0 \ll vR$ the difference between the $T$-matrix and $U$ is negligible. Furthermore the logarithm in Eq.~(\ref{W_AA}) can be expanded yielding the perturbative expression, Eq. (1),
\begin{eqnarray}
\label{W_AA_pert} W_{\it \! AA}({\bf R})= -\frac{ U^2 A_0^2}{\pi^2 v^4} \cos^2\theta_{\it \! AA}\! \!
\int\limits_{-\infty}^\infty d\omega \omega^2  K^2_0\left(\frac{|\omega| R}{v}\right)\nonumber\\ =-\frac{1}{16\pi} \frac{U^2 \! A_0^2}{v R^3}\cos^2\!{\theta_{\! AA}}.~~~~
\end{eqnarray}
In the strong impurity limit $UA_0 \ll vR$ the $T$-matrix is independent of $U$,
\begin{equation}
\label{T-matrix_strong}
T_{i\omega} =\frac{-i \pi v^2}{\omega A_0 \ln{\left(\frac{t}{|\omega|}\right)}}.
\end{equation}
The leading contribution to the frequency integral still comes from $|\omega|\sim v/R$ where $K_0 \sim 1$. Provided that $\ln(R/a) \gg 1$ the expansion to the lowest order in $T_{i\omega}$ remains legitimate and we obtain Eq. (3),
\begin{equation}
\label{W_AA_strong} W_{\it \! AA}({\bf R})= -\frac{\cos^2\theta_{\it \! AA} }{\ln^2{(R/a)}}
\int\limits_{-\infty}^\infty \frac{d\omega}{\pi}  K^2_0\left(\frac{|\omega| R}{v}\right)=\frac{\pi v \cos^2\!{\theta_{\! AA}}}{2R\ln^2\!{(R/a)}}.
\end{equation}

\section{{\it AB}-configuration of two impurities}
When both impurities are residing on the same sublattice, using Eq.~(\ref{G_AA}) we find that the interaction energy is given by
\begin{eqnarray}
\label{W_AB} W_{\it \! AB}({\bf R})= -2
\int\limits_{-\infty}^\infty\frac{d\omega}{2\pi}
\ln{\left(1-T^2_{i\omega}G_{i\omega}({\bf R},0)G_{i\omega}(0,{\bf R})  \right)}=\nonumber\\ -2
\int\limits_{-\infty}^\infty\frac{d\omega}{2\pi}
\ln{\left(1-T^2_{i\omega} \frac{\omega^2 A_0^2}{\pi^2 v^4} K^2_1\left({|\omega| R}/{v}\right) \sin^2\theta_{\it \! AB} \right)}.~~~
\end{eqnarray}
In the week impurity limit (Eq. (2) of the paper),
\begin{eqnarray}
\label{W_AB_pert} W_{\it \! AB}({\bf R})= \frac{ U^2 A_0^2}{\pi v^4} \sin^2\theta_{\it \! AB}
\int\limits_{-\infty}^\infty d\omega \omega^2  K^2_1\left(\frac{|\omega| R}{v}\right) \nonumber\\ =\frac{3}{16\pi} \frac{U^2 \! A_0^2}{v R^3}\sin^2\theta_{\it \! AB}.
\end{eqnarray}
In the strong impurity limit
\begin{eqnarray}
\label{W_AB_strong} W_{\it \! AB}({\bf R})= -2
\int\limits_{-\infty}^\infty\frac{d\omega}{2\pi}
\ln{\left(1+\frac{K^2_1\left({|\omega| R}/{v}\right) \sin^2\theta_{\it \! AB}}{\ln^2\!{(R/a)}} \right)} \nonumber\\ =-\frac{2v}{\pi R}
\int\limits_{0}^\infty dy
\ln{[1+z^2 K^2_1(y)]},~~~
\end{eqnarray}
where $z=|\sin\theta_{\it \! AB}|/\ln\!{(R/a)}\ll 1$. In the last integral we can write,
\begin{eqnarray}
\label{first}
\int\limits_{0}^\infty dy
\ln{\left(1+z^2 K^2_1(y) \right)}= \int\limits_{0}^\infty dy
\ln{\left(1+\frac{z^2}{y^2} \right)}\nonumber\\+\int\limits_{0}^\infty dy
\ln{\left(\frac{1+z^2 K^2_1(y)}{1+{z^2}/{y^2}} \right)}.
\end{eqnarray}
The first integral in the last expression originates from $y \sim z \ll 1$ and is equal to $\pi z$. In the second integral,
the main contribution comes from $y\sim 1$, since the leading singularity has been subtracted. Thus, for $z \ll 1$ the logarithms can now we expanded to the linear order in $z^2$ to give
\begin{equation}
\label{second}
\int\limits_{0}^\infty dy
\ln{\left(\frac{1+z^2 K^2_1(y)}{1+{z^2}/{y^2}} \right)}=z^2 \int\limits_{0}^\infty dy~
[K^2_1(y) -1/y^{2}]= -\frac{\pi^2 z^2}{4}.
\end{equation}
as a result,
\begin{eqnarray}
\label{W_AB_strong2} W_{\it \! AB}({\bf R})= -\frac{2v}{\pi R}
\int\limits_{0}^\infty dy
\ln{[1+z^2 K^2_1(y)]}\nonumber\\ =-\frac{2v}{\pi R} \left(\pi z -\frac{\pi^2 z^2}{4}+O(z^3)\right),
\end{eqnarray}
which reproduces Eq.~(4).


\begin{references}

\bibitem{CN} A.~H.~Castro Neto,
F.~Guinea, N.~M.~R.~Peres, K.~S.~Novoselov, and A.~K.~Geim, Rev.
Mod. Phys. 81, 109 (2009)


\bibitem{CF} V. V. Cheianov and V. I. Fal'ko,
Phys. Rev. Lett. {\bf 97}, 226801 (2006).

\bibitem{KCM} A. B. Kuzmenko, I. Crassee, D. van der Marel, P. Blake, and K. S.
Novoselov, Phys. Rev. B {\bf 80}, 165406 (2009).

\bibitem{ZTG} Y. Zhang, T.-T. Tang, C. Girit, Z. Hao, M. C. Martin, A. Zettl,
M.F. Crommie, Y. R. Shen, and F. Wang, Nature {\bf 459}, 820 (2009).

\bibitem{MLS} K. F. Mak, C. H. Lui, J. Shan, and T. F. Heinz, Phys. Rev.
Lett. {\bf 102}, 256405 (2009).

\bibitem{GKB} G. Giovannetti, P. A. Khomyakov, G. Brocks, P. J. Kelly, and J. van den Brink,
Phys. Rev. B {\bf 76}, 073103 (2007).

\bibitem{NWM} Z. H. Ni, H. M. Wang, Y. Ma, J. Kasim, Y. H. Wu, and Z. X.
Shen, ACS Nano {\bf 2}, 1033 (2008).

\bibitem{NYL} Z. H. Ni, T. Yu, Y. H. Lu, Y. Y. Wang, Y. P. Feng, and Z. X. Shen,
ACS Nano {\bf 2}, 2301 (2008).

\bibitem{SN} P. Shemella and S. K. Nayak, Appl. Phys. Lett. {\bf 94}, 032101
(2009).

\bibitem{KZJ} K. S. Kim, Y. Zhao, H. Jang, S. Y. Lee, J. M. Kim, K. S. Kim, J.
H. Ahn, P. Kim, J. Y. Choi, and B. H. Hong, Nature (London) {\bf 457}, 706 (2009).

\bibitem{PCP} V. M. Pereira, A. H. Castro Neto, and N. M. R. Peres, Phys. Rev.
B {\bf 80}, 045401 (2009).

\bibitem{CCC} G. Cocco, E. Cadelano, and L. Colombo, Phys. Rev. B {\bf 81}, 241412(R) (2010).

\bibitem{E} M. Ezawa, Phys. Rev. B {\bf 73}, 045432 (2006).

\bibitem{SCL} Y.-W. Son, M. L. Cohen, and S. G. Louie, Phys. Rev. Lett. {\bf 97}, 216803 (2006).

\bibitem{HOZ} M. Y. Han, B. \"Ozyilmaz, Y. Zhang, and P. Kim, Phys. Rev. Lett.
{\bf 98}, 206805 (2007).

\bibitem{KM} C. L. Kane and E. J. Mele, Phys. Rev. Lett. {\bf 95}, 146802 (2005).

\bibitem{QYF} Zh. Qiao, Sh. A. Yang, W. Feng, W.-K. Tse, J. Ding, Y. Yao, J. Wang, and Q. Niu, Phys. Rev. B {\bf 82}, 161414(R) (2010).

\bibitem{WHA}  C. Weeks, J. Hu, J. Alicea, M. Franz, and R. Wu, Phys. Rev. X {\bf 1}, 021001 (2011).

\bibitem{BME}  A. Bostwick, J. L. McChesney, K. V. Emtsev, T. Seyller, K. Horn, S. D. Kevan, and E. Rotenberg, Phys. Rev. Lett. {\bf 103}, 056404 (2009).

\bibitem{ENM} D. C. Elias, R. R. Nair, T. M. G. Mohiuddin, S. V. Morozov, P. Blake, M. P. Halsall, A. C. Ferrari, D. W. Boukhvalov, M. I. Katsnelson, A. K. Geim, K. S. Novoselov, Science {\bf 323}, 610 (2009).

\bibitem{BJN} R. Balog, B. J\o rgensen, L. Nilsson, M. Andersen, E. Rienks, M.
Bianchi, M. Fanetti, E. L\o gsgaard, A. Baraldi, S. Lizzit, Z. Sljivancanin,
F. Besenbacher, B. Hammer, T. G. Pedersen, P. Hofmann,
and L. Hornek�, Nature (London) {\bf 9}, 315 (2010).

\bibitem{DQF} J. Ding, Zh. Qiao, W. Feng, Y. Yao, and Q. Niu, Phys. Rev. B {\bf 84}, 195444 (2011).

\bibitem{CSA} V. V. Cheianov, O. Syljuasen, B. L. Altshuler, and V.I. Falko, Europhys. Lett. {\bf 89}, 56003 (2010).

\bibitem{ASL} D. A. Abanin, A. V. Shytov, and L. S. Levitov,
Phys. Rev. Lett. {\bf 105}, 086802 (2010).

\bibitem{KCA} S. Kopylov, V. Cheianov, B. L. Altshuler, and V.I. Fal'ko, Phys. Rev. B {\bf 83}, 201401(R) (2011).

\bibitem{MT}  V. Mostepanenko and N. Trunov, {\it The Casimir Effect and Its
Applications} (Clarendon, Oxford, 1997).

\bibitem{Friedel} J. Friedel, Philos. Mag. {\bf 43}, 153 (1952).

\bibitem{LK} K. H. Lau and W. Kohn, Surf. Sci. {\bf 75}, 69 (1978).

\bibitem{MF} E. McCann and V. I. Fal'ko, Phys. Rev. Lett. {\bf 96}, 086805 (2006).

\bibitem{S}  
    \'{A}. B\'{a}csi and A. Virosztek, Phys. Rev. B {\bf 82}, 193405 (2010).

\bibitem{MilT} The continuous limit of Eqs.~(\ref{first_AA}-\ref{first_AB}) was considered in A. I. Milstein and I. S. Terekhov, Phys. Rev. B {\bf 81}, 125419 (2010); V. V. Mkhitaryan and E. G. Mishchenko, Phys. Rev. B {\bf 86}, 115442 (2012).

\bibitem{KK} Such resummation is easily performed exactly  in case of two point-like objects: O. Kenneth and I. Klich, Phys. Rev. Lett. {\bf 97}, 160401 (2006); T. Emig, N. Graham, R. L. Jaffe, and M. Kardar, Phys. Rev. Lett. {\bf 99}, 170403 (2007).

\bibitem{WKL} T. O. Wehling, M. I. Katsnelson, A. I. Lichtenstein, Chem. Phys. Lett {\bf 476}, 125 (2009).

\bibitem{SAL} A. V. Shytov, D. A. Abanin, and L. S. Levitov,
Phys. Rev. Lett. {\bf 103}, 016806 (2009).


\bibitem{SJR}  D. Solenov, C. Junkermeier, T. L. Reinecke, and K. A. Velizhanin, Phys. Rev. Lett. {\bf 111}, 115502 (2013).

\bibitem{shytov} This is the same value as used in Ref.~\cite{SAL}: A. V. Shytov, private communication.

\bibitem{LLQM} L. D. Landau and E. M. Lifshitz, {\it Quantum Mechanics} (Pergamon Press, Oxford, 1976).



\end{references}
\end{document}